\documentclass[twocolumn,american,english]{revtex4}
\usepackage[T1]{fontenc}
\usepackage[latin9]{inputenc}
\usepackage{color}
\usepackage{amstext}
\usepackage{graphicx}

\makeatletter
\@ifundefined{textcolor}{}
{%
 \definecolor{BLACK}{gray}{0}
 \definecolor{WHITE}{gray}{1}
 \definecolor{RED}{rgb}{1,0,0}
 \definecolor{GREEN}{rgb}{0,1,0}
 \definecolor{BLUE}{rgb}{0,0,1}
 \definecolor{CYAN}{cmyk}{1,0,0,0}
 \definecolor{MAGENTA}{cmyk}{0,1,0,0}
 \definecolor{YELLOW}{cmyk}{0,0,1,0}
}

\makeatother

\usepackage{babel}
\begin{document}

\title{Phase Dilemma in Natural Orbital Functional Theory from the N-representability
Perspective}

\author{Ion Mitxelena$^{1,2}$, Mauricio Rodríguez-Mayorga$^{2,3}$, Mario
Piris$^{1,2,4}$}
\email{mario.piris@ehu.eus}

\address{$^{1}$Kimika Fakultatea, Euskal Herriko Unibertsitatea (UPV/EHU),
P.K. 1072, 20080 Donostia, Euskadi, Spain.}

\address{$^{2}$Donostia International Physics Center (DIPC), 20018 Donostia,
Euskadi, Spain.}

\address{$^{3}$Institut de Química Computacional i Catàlisi, 17003 Girona,
Catalonia, Spain.}

\address{$^{4}$IKERBASQUE, Basque Foundation for Science, 48013 Bilbao, Euskadi,
Spain.}
\begin{abstract}
Any rigorous approach to first-order reduced density ($\Gamma$) matrix
functional theory faces the phase dilemma, that is, having to deal
with a large number of possible combinations of signs in terms of
the electron-electron interaction energy. This problem was discovered
by reducing a ground-state energy generated from an approximate N-particle
wavefunction into a functional of $\Gamma$, known as the top-down
method. Here, we show that the phase dilemma also appears in the bottom-up
method, in which the functional $E\left[\Gamma\right]$ is generated
by progressive inclusion of N-representability conditions on the reconstructed
two-particle reduced density matrix. It is shown that an adequate
choice of signs is essential to accurately describe model systems
with strong non-dynamic (static) electron correlation, specifically,
the one-dimensional Hubbard model with periodic boundary conditions
and hydrogen rings. For the latter, the Piris natural orbital functional
7 (PNOF7), with phases equal to -1 for the inter-pair energy terms
containing the exchange-time-inversion integrals, agrees with exact
diagonalization results.
\end{abstract}
\maketitle
The first-order reduced density matrix ($\Gamma$) functional theory,
that is, the theory where the ground-state energy ($E$) is represented
in terms of $\Gamma$, has emerged in recent years as a promising
method to study strongly correlated materials. The seminal article
of Gilbert \cite{Gilbert1975} on the existence of the functional
along with the works of Donnelly and Parr \cite{Donnelly1978}, Levy
\cite{Levy1979} and Valone \cite{Valone1980} laid the foundations,
however, the computational schemes based on these exact formulations
are several times more expensive than solving directly the Schrödinger
equation, so practical applications require a different approach for
$E\left[\Gamma\right]$.

In 1967 \cite{Rosina1967}, Rosina demonstrated that there is a one-to-one
mapping from the ground-state two-particle reduced density matrix
(D) to the N-particle wavefunction in the case of a Hamiltonian with
at most two-body interactions. Taking advantage of the Rosina's theorem,
the existence theorem of Gilbert implicitly establishes a one-to-one
correspondence between the ground-state D and $\Gamma$, therefore,
the functional $E\left[\Gamma\right]$ must match the exact well-known
functional $E\left[\mathrm{D}\right]$. It should be noted that the
unknown functional in a $\Gamma$-based theory is the electron-electron
potential energy ($V_{ee}$) since the rest of the Hamiltonian is
actually a single-particle operator. Unfortunately, the exact reconstruction
$V_{ee}\left[\Gamma\right]$ has been an unattainable goal so far,
and we have to settle for making approximations. The typical approach
is to employ the exact energy expression $E\left[\mathrm{D}\right]$
but using solely a reconstruction functional $\mathrm{D}\left[\Gamma\right]$.

Approximating the energy functional in that way implies that theorems
obtained for the exact functional $E\left[\Gamma\right]$ are no longer
valid, since an approximate functional still depends on D \cite{Donnelly1979}.
An undesired implication is that the functional N-representability
problem arises \cite{Piris2010a,Ludena2013,Piris2018}, that is, we
have to comply the requirement that D reconstructed in terms of $\Gamma$
must satisfy the same N-representability conditions as those imposed
on unreconstructed two-matrices to ensure a physical value of the
approximate ground-state energy; otherwise, there will not be an N-electron
system with an energy value $E\left[\mathrm{D}\right]$.

In general, $\Gamma$-based functionals have been proposed using reasonable
heuristic or physical arguments \cite{Pernal2016}, so that most of
the approximate functionals currently in use are not N-representable,
and that is why energy is often obtained far below true energy. It
has been assumed that there is no N-representability problem of the
functional, since it is believed that only N-representable conditions
\cite{Coleman1963} on $\Gamma$ are sufficient, but the latter is
only true for the exact reconstruction of $V_{ee}\left[\Gamma\right]$.
The ensemble N-representability constraints for acceptable $\Gamma$
are easy to implement, but are insufficient to guarantee that the
reconstructed D is N-representable, and thereby the approximate functional
either. To date, only the functionals proposed by Piris and coworkers
\cite{Piris2006,Piris2013b,Piris2014c,Piris2017} relies on the reconstruction
of D subject to ensemble N-representability conditions.

One issue related to the functional N-representability is that approximate
N-representable functionals are not invariant with respect to a unitary
transformation of the orbitals. The fact is that apart from the simple
Hartree-Fock (HF) approximation, none of the known approximate functionals
are explicitly given in terms of $\Gamma$, including the familiar
functional which accurately describes two-electron closed-shell systems
\cite{Lowdin1955d,Goedecker2000}. It is worth noting that there are
functionals \cite{Muller1984,Csanyi2000,Marques2008,Sharma2008} that
seem to depend properly on $\Gamma$. However, these functionals violate
the antisymmetric requirement for D \cite{Herbert2003}, consequently
none of these functionals affords an N-representable two-matrix, nor
can they reproduce the simplest two-electron case. Extensive N-representability
violations have been recently reported \cite{Rodriguez-Mayorga2017}
for this kind of functionals.

Approximations for $E\left[\Gamma\right]$ can be obtained essentially
using two methods, namely, the top-down and bottom-up methods \cite{Ludena2013,Piris2013a}.
The top-down method consists in the reduction of an N-particle ground-state
energy generated from an approximate wavefunction into a functional
of $\Gamma$, whereas, in the bottom-up method $E\left[\Gamma\right]$
is generated by progressive inclusion of N-representability conditions
\cite{Mazziotti2007a} on the reconstructed $\mathrm{D}\left[\Gamma\right]$.

The use of the top-down method with a parametrization of coefficients
in a configuration interaction (CI) expansions reveals a very serious
bottleneck affecting any rigorous approach to $E\left[\Gamma\right]$,
namely the phase dilemma that stems from the necessity to carry out
minimization over a large number of possible combinations of CI coefficient
signs \cite{Cioslowski2004}. As expected, the phase dilemma also
appears when the bottom-up method is used, i.e., we have to deal with
a large number of possible combinations of signs in terms of the electron-electron
interaction energy.

In the next section, we analyze how the phase dilemma arises when
applying N-representabilty conditions to the reconstructed $\mathrm{D}\left[\Gamma\right]$.
In sections \ref{sec:The-Hubbard-Model} and \ref{sec:Hydrogen-rings},
we demonstrate that a suitable choice of signs is essential to describe
accurately model systems with strong non-dynamic (static) electron
correlation. This leads us to the formulation of the Piris natural
orbital functional 7 (PNOF7) with phases equal to -1 for the inter-pair
energy terms containing the exchange-time-inversion integrals, which
captures the electron correlation energy close to the exact diagonalization
values.

\section{Natural Orbital Functional Theory and N-representability}

The present-day functionals are constructed in the basis where $\Gamma$
is diagonal, which is the definition of a natural orbital functional
(NOF) \cite{Piris2007,Piris2014a}. In this context, the natural orbitals
(NOs) are the orbitals that diagonalize the one-matrix corresponding
to an approximate ground-state energy, so it is more appropriate to
speak of a NOF rather than a functional of $\Gamma$ due to the explicit
dependence on D mentioned above for approximate functionals. 

Accordingly, the ground-state electronic energy is given in terms
of the NOs and their occupation numbers (ONs), namely,
\begin{equation}
E=\sum\limits _{i}n_{i}\mathcal{H}_{ii}+\sum\limits _{ijkl}D[n_{i},n_{j},n_{k},n_{l}]<kl|ij>\label{ENOF}
\end{equation}
where $\mathcal{H}_{ii}$ denotes the diagonal elements of the core-Hamiltonian,
$<kl|ij>$ are the matrix elements of the two-particle interaction,
and $D[n_{i},n_{j},n_{k},n_{l}]$ represents the reconstructed D from
the ONs. Restrictions on the ONs to the range $0\leq n_{i}\leq1$,
also known as Pauli constraints, represent the necessary and sufficient
conditions for ensemble N-representability of $\Gamma$ \cite{Coleman1963}
under the normalization condition $\sum_{i}n_{i}=\mathrm{N}$. 

On this respect, it is worth noting that we focus on the N-representability
problem for statistical ensembles. Conditions named generalized Pauli
constraints have been obtained \cite{Klyachko2006,Altunbulak2008}
for pure-state N-representability of $\Gamma$. However, the number
of these conditions increases dramatically with the number of NOs,
so it becomes quite difficult to handle them in practical implementations
\cite{Theophilou2015}. Anyway, in order to guarantee the pure-state
N-representability conditions in the minimization of $E\left[\Gamma\right]$
only Pauli constraints are necessary if the functional is the appropriate
one \cite{Valone1980,Nguyen-Dang1985}. Indeed, if the approximate
NOF is pure-state N-representable, i.e., it is obtained from the reconstruction
of a pure-state N-representable two-matrix, then contraction of D
will always lead to a pure-state N-representable one-matrix.

It is clear that the construction of an N\textit{-}representable functional
given by (\ref{ENOF}) is related to the N-representability problem
of $D[n_{i},n_{j},n_{k},n_{l}]$. The use of ensemble N-representability
conditions \cite{Mazziotti2012} for generating a reconstruction functional
was proposed in Ref. \cite{Piris2006}. This particular reconstruction
is based on the introduction of two auxiliary matrices $\mathbf{\triangle}$
and $\Pi$ expressed in terms of the ONs to reconstruct the cumulant
part of D \cite{Mazziotti1998}. In this work, we address only singlet
states and adopt a restricted spin theory, so that energy (\ref{ENOF})
becomes
\begin{equation}
\begin{array}{c}
E=2\sum\limits _{p}n_{p}\mathcal{H}_{pp}+\sum\limits _{qp}\Pi_{qp}\mathcal{L}_{pq}\qquad\qquad\\
+\sum\limits _{qp}\left(n_{q}n_{p}-\Delta_{qp}\right)\left(2\mathcal{J}_{pq}-\mathcal{K}_{pq}\right)
\end{array}\label{PNOF}
\end{equation}
where $\mathcal{J}_{pq}$, $\mathcal{K}_{pq}$, and $\mathcal{L}_{pq}$
are the direct, exchange, and exchange-time-inversion integrals \cite{Piris1999}.
Appropriate forms of matrices $\Delta$ and $\Pi$ lead to different
implementations known in the literature as PNOFi (i=1-7) \cite{Piris2006,Piris2013b,Piris2014c,Piris2017}.
Remarkable is the case of PNOF5 \cite{Piris2011,Piris2013e} which
turned out to be strictly pure N-representable \cite{Pernal2013,Piris2013c}.

The conservation of the total spin allows us to derive the diagonal
elements $\Delta_{pp}=n_{p}^{2}$ and $\Pi_{pp}=n_{p}$ \cite{Piris2009}.
The N-representability $D$ and $Q$ conditions of the two-matrix
impose the following inequalities on the off-diagonal elements of
$\Delta$ \cite{Piris2006},
\begin{equation}
\begin{array}{c}
\Delta_{qp}\leq n_{q}n_{p}\end{array},\qquad\Delta_{qp}\leq h_{q}h_{p}\label{DQ_cond}
\end{equation}
whereas to fulfill the $G$ condition, the elements of the $\Pi$-matrix
must satisfy the constraint \cite{Piris2010a}
\begin{equation}
\Pi_{qp}^{2}\leq\left(n_{q}h_{p}+\Delta_{qp}\right)\left(h_{q}n_{p}+\Delta_{qp}\right)\label{G_cond}
\end{equation}
Here, $h_{p}$ denotes the hole $1-n_{p}$. Notice that for singlets
the total hole for a given spatial orbital $p$ is $2h_{p}$.

For a given approximation of $\Delta_{qp}$ that satisfies the inequalities
of Eq. (\ref{DQ_cond}), it is evident that the modulus of $\Pi$
matrix elements is determined from Eq. (\ref{G_cond}) assuming the
equality, however, there is not any hint to determine the sign of
$\Pi_{qp}$. Consequently, a large number of possible combinations
of these signs looms up for those terms containing exchange-time-inversion
integrals $\mathcal{L}_{pq}$ in Eq. (\ref{PNOF}). We need to solve
this phase dilemma to propose a NOF, that is, make an adequate selection
of the $\Pi_{qp}$ signs.

We now focus on the simplest case of two electrons. Fortunately, an
accurate NOF is well-known for this system from the exact wavefunction
\cite{Lowdin1955d} assuming that all amplitudes, with the exception
of the first one, are negative if the first amplitude is chosen to
be positive \cite{Goedecker2000}. The two-electron singlet energy
reads as
\begin{equation}
\begin{array}{c}
E\left(2e^{-}\right)=2\sum\limits _{p=1}^{\infty}n_{p}\mathcal{H}_{pp}+n_{1}\mathcal{L}_{11}\\
+\sum\limits _{p,q=2}^{\infty}\sqrt{n_{q}n_{p}}\mathcal{L}_{pq}-2{\displaystyle \sum_{p=2}^{\infty}\sqrt{n_{1}n_{p}}\mathcal{L}_{p1}}
\end{array}\label{2e}
\end{equation}

It is worth mentioning that in some stretched two-electron molecules,
small contributions to energy may have opposite signs to those adopted
in Eq. (\ref{2e}) \cite{Davidson1976,Sheng2013}. A recent study
\cite{Rodriguez-Mayorga2017} on the two-electron Harmonium atom reveals
similar small deviations in the high-correlation regime. Nevertheless,
the convention of signs adopted in Eq. (\ref{2e}) provides very accurate
results for almost all correlation regimes in two-electron systems,
including those with strong non-dynamic correlation.

The requirement that for any two-electron singlet system the NOF (\ref{PNOF})
yields the accurate expression (\ref{2e}), together with the cumulant
sum rules, and the N-representability conditions (\ref{DQ_cond})
and (\ref{G_cond}), imply that $\Delta_{qp}=n_{q}n_{p}$ and $\left|\Pi_{qp}\right|=\sqrt{n_{q}n_{p}}$,
respectively \cite{Piris2010a}. Furthermore, the phase factor of
$\Pi_{qp}$ is $+1$ if $q,p\in\left(1,\infty\right)$, and -1 otherwise.

For systems with N>2, the generalization of these expressions for
auxiliary matrices $\mathbf{\triangle}$ and $\Pi$ leads to the independent
pair model (PNOF5) \cite{Piris2011,Piris2013e}:
\begin{equation}
\begin{array}{c}
\Delta_{qp}=n_{p}^{2}\delta_{qp}+n_{q}n_{p}\left(1-\delta_{qp}\right)\delta_{q\Omega_{g}}\delta_{p\Omega_{g}}\\
\\
\Pi_{qp}=n_{p}\delta_{qp}+\Pi_{qp}^{g}\left(1-\delta_{qp}\right)\delta_{q\Omega_{g}}\delta_{p\Omega_{g}}\\
\\
\Pi_{qp}^{g}=\left\{ \begin{array}{cc}
-\sqrt{n_{q}n_{p}}\,, & p=g\textrm{ or }q=g\\
+\sqrt{n_{q}n_{p}}\,, & p,q>\mathrm{N}/2
\end{array}\right.\\
\\
\delta_{q\Omega_{g}}=\left\{ \begin{array}{cc}
1\,, & q\in\Omega_{g}\\
0\,, & q\notin\Omega_{g}
\end{array}\right.;\quad g=1,2,\ldots,\mathrm{N}/2
\end{array}\label{PNOF5}
\end{equation}
where we have divided the orbital space $\Omega$ into N/2 mutually
disjoint subspaces $\Omega{}_{g}$, so each orbital belongs only to
one subspace. Each subspace contains one orbital $g$ below the Fermi
level (N/2), and $\mathrm{N}_{g}$ orbitals above it, which is reflected
in additional sum rules for the ONs:
\begin{equation}
\sum_{p\in\Omega_{g}}n_{p}=1;\quad g=1,2,\ldots,\mathrm{N}/2\label{sumrule_n}
\end{equation}
Taking into account the spin, each subspace contains solely an electron
pair, and the normalization condition for $\Gamma$ ($2\sum_{p}n_{p}=\mathrm{N}$)
is automatically fulfilled.

The energy (\ref{PNOF}) of PNOF5 can be then conveniently written
as
\begin{equation}
\begin{array}{c}
E=\sum\limits _{g=1}^{\mathrm{N}/2}E_{g}+\sum\limits _{f\neq g}^{\mathrm{N}/2}E_{fg}\\
\\
E_{g}=\sum\limits _{p\in\Omega_{g}}n_{p}\left(2\mathcal{H}_{pp}+\mathcal{J}_{pp}\right)+\sum\limits _{p,q\in\Omega_{g},p\neq q}\Pi_{qp}^{g}\mathcal{L}_{pq}\\
\\
E_{fg}=\sum\limits _{q\in\Omega_{f}}\sum\limits _{p\in\Omega_{g}}n_{q}n_{p}\left(2\mathcal{J}_{pq}-\mathcal{K}_{pq}\right)
\end{array}\label{EPNOF5}
\end{equation}
The first term of the energy draws the system as independent N/2 electron
pairs, whereas the second term contains the contribution to the HF
mean-field of the electrons belonging to different pairs.

To go beyond the independent-pair approximation, let's maintain $\mathrm{\Delta_{\mathit{qp}}=0}$
and consider nonzero the $\Pi$-elements between orbitals belonging
to different subspaces \cite{Piris2017}. From Eq. (\ref{G_cond}),
note that provided the $\Delta_{qp}$ vanishes, $\left|\Pi_{qp}\right|\leq\Phi_{q}\Phi_{p}$
with $\Phi_{q}=\sqrt{n_{q}h_{q}}$. Assuming equality once again,
the sign of $\Pi_{qp}$ remains undetermined, so there is a large
number of possible combinations of signs that affect now the inter-pair
interactions. Again, we need to solve this phase dilemma to propose
a NOF, but this time we need to make a proper selection of the $\Pi_{qp}$
signs for orbitals $q$ and $p$ belonging to different pairs (subspaces).
In contrast to the intra-pair interactions, there is no indication
to determine the phase factor for the inter-pair $\Pi_{qp}$.

Recently \cite{Piris2017}, the generalization of the sign convention
adopted for $\Pi_{qp}^{g}$ in Eq. (\ref{PNOF5}), namely $\Pi_{qp}^{\Phi}=\Phi_{q}\Phi_{p}$
if $q,p>\mathrm{N}/2$, and $\Pi_{qp}^{\Phi}=-\Phi_{q}\Phi_{p}$ otherwise,
led to a new functional denoted as PNOF7. The resulting energy is
\begin{equation}
\begin{array}{c}
E=\sum\limits _{g=1}^{\mathrm{N}/2}E_{g}+\sum\limits _{f\neq g}^{\mathrm{N}/2}E_{fg}\\
\\
E_{g}=\sum\limits _{p\in\Omega_{g}}n_{p}\left(2\mathcal{H}_{pp}+\mathcal{J}_{pp}\right)+\sum\limits _{q,p\in\Omega_{g},q\neq p}\Pi_{qp}^{g}\mathcal{L}_{pq}\\
\\
E_{fg}=\sum\limits _{p\in\Omega_{f}}\sum\limits _{q\in\Omega_{g}}\left[n_{q}n_{p}\left(2\mathcal{J}_{pq}-\mathcal{K}_{pq}\right)+\Pi_{qp}^{\Phi}\mathcal{L}_{pq}\right]
\end{array}\label{PNOF7}
\end{equation}
It is obvious that a possible election that favors decreasing of the
energy (\ref{PNOF7}) is to consider all the phase factors negative,
i.e., $\Pi_{qp}^{\Phi}=-\Phi_{q}\Phi_{p}$. From now on we will denote
by PNOF7(+) the functional that considers +1 the phase factors of
$\Pi_{qp}^{\Phi}$ for $q,p>\mathrm{N}/2$, whereas PNOF7(-) will
be employed to denote the NOF (\ref{PNOF7}) with these phases equal
to -1.

Since we do not have an accurate functional like (\ref{2e}) that
helps us to determine which is the best combination of signs for $\Pi_{qp}^{\Phi}$,
in the following sections, we analyze several examples with strong
non-dynamic (static) electron correlation in order to make a proper
phase choice after comparing with exact diagonalization calculations.

\section{\label{sec:The-Hubbard-Model}Hubbard model}

The Hubbard model is an ideal candidate for the study of electron
correlation due to its conceptual simplicity. The corresponding one-dimensional
(1D) Hamiltonian reads as \cite{hubbard} 
\begin{equation}
\begin{array}{c}
{\displaystyle H=-t\sum_{<\mu,\upsilon>}\left(c_{\mu,\sigma}^{\dagger}c_{\upsilon,\sigma}+c_{\mu,\sigma}c_{\upsilon,\sigma}^{\dagger}\right)}\\
\\
{\displaystyle +U\sum_{\mu}n_{\mu,\alpha}n_{\mu,\beta}}
\end{array}\label{hubbard-equation}
\end{equation}
where greek indices $\mu$ and $\upsilon$ denote sites of the model,
$<\mu,\upsilon>$ indicates only near-neighbors interactions, $t>0$
is the hopping parameter, $\sigma=\alpha,\beta$; $U$ is the on-site
inter-electron repulsion parameter, and $n_{\mu,\sigma}=c_{\mu,\sigma}^{\dagger}c_{\mu,\sigma}$
where $c_{\mu,\sigma}^{\dagger}\left(c_{\mu,\sigma}\right)$ corresponds
to fermionic creation(annihilation) operator. It is known that the
HF approximation retrieves the exact solution for the 1D Hubbard model
at half-filling if $U=0$, where, in the site basis, all the possible
states $|--\rangle$, $|-\uparrow\rangle$, $|-\downarrow\rangle$,
and $|\downarrow\uparrow\rangle$ have the same weight. Conversely,
in the $U/t\rightarrow\infty$ limit the singly occupied states $|-\uparrow\rangle$
and $|-\downarrow\rangle$ appear uniquely, so the antiferromagnetic
scheme is recovered and the model becomes equivalent to the spin-1/2
Heisenberg model.

The performance of commonly used NOF approximations in the 1D Hubbard
model with periodic boundary conditions has been recently studied
\cite{mitxelena-hubbard,mitxelena-erratum}, showing that the here
presented PNOF7(+) is in good agreement with exact results for the
Hubbard model at half-filling. Nevertheless, the amount of electron
correlation recovered by PNOF7(+) for large systems is slightly less
than for small systems. Since the NOF theory is a promising approach
for large many-body systems, it is crucial to develop approximations
that do not deteriorate as the size of the system increases. In the
following, we show that a proper choice of inter-pair interaction
signs prevents the accumulation of errors as the number of electron
pairs in the system gets larger.

\begin{figure}
\includegraphics[scale=0.68]{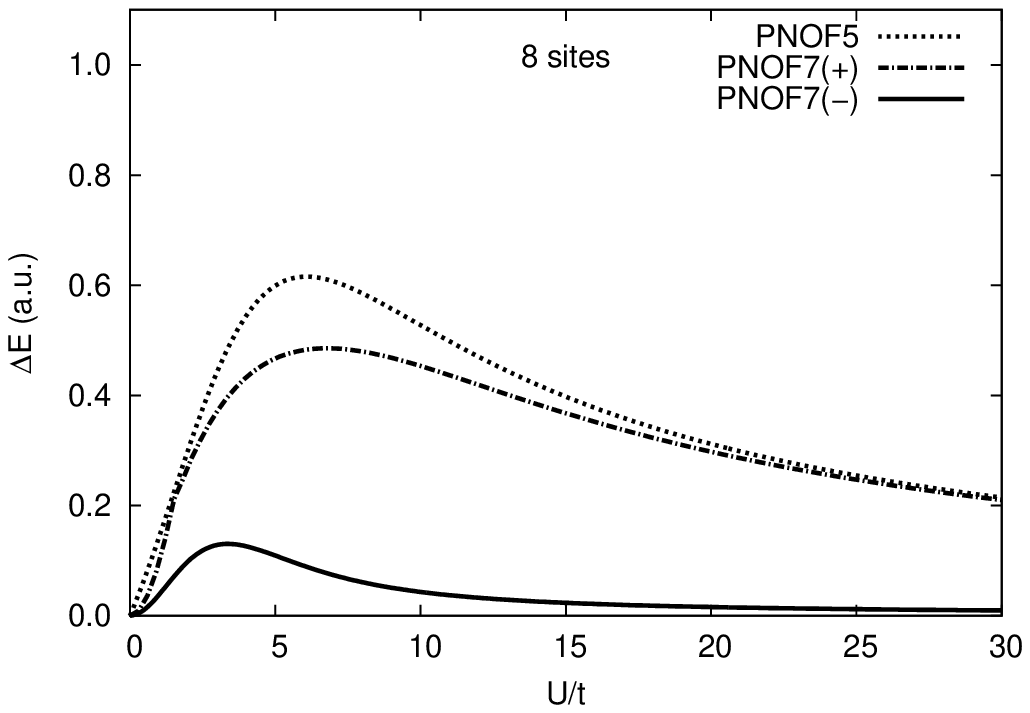}

\includegraphics[scale=0.68]{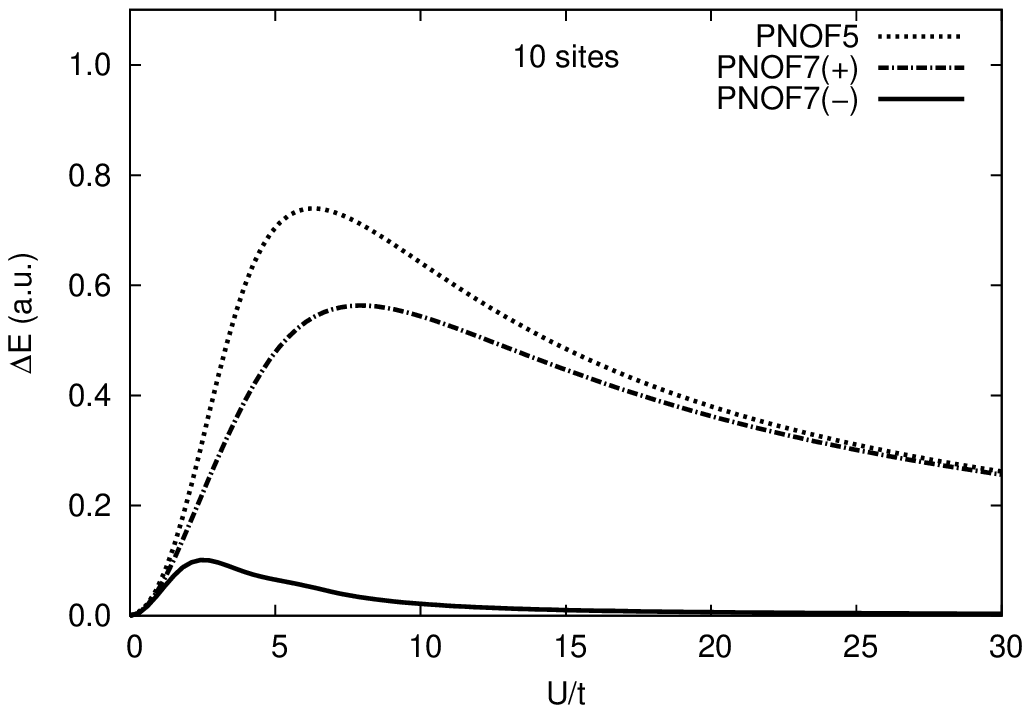}

\includegraphics[scale=0.68]{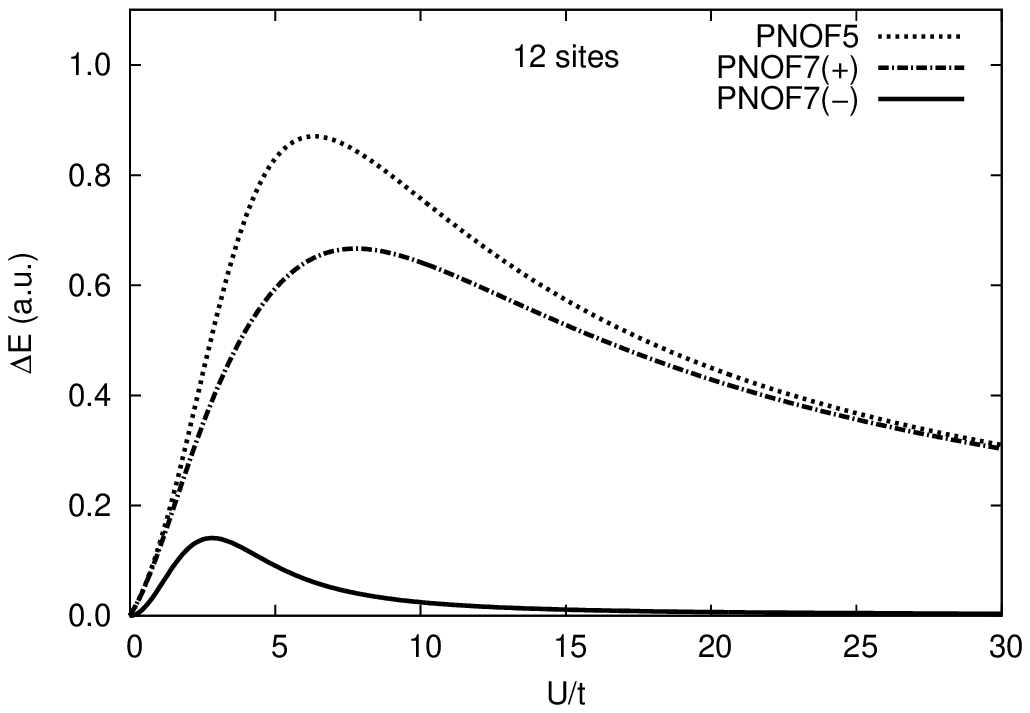}

\includegraphics[scale=0.68]{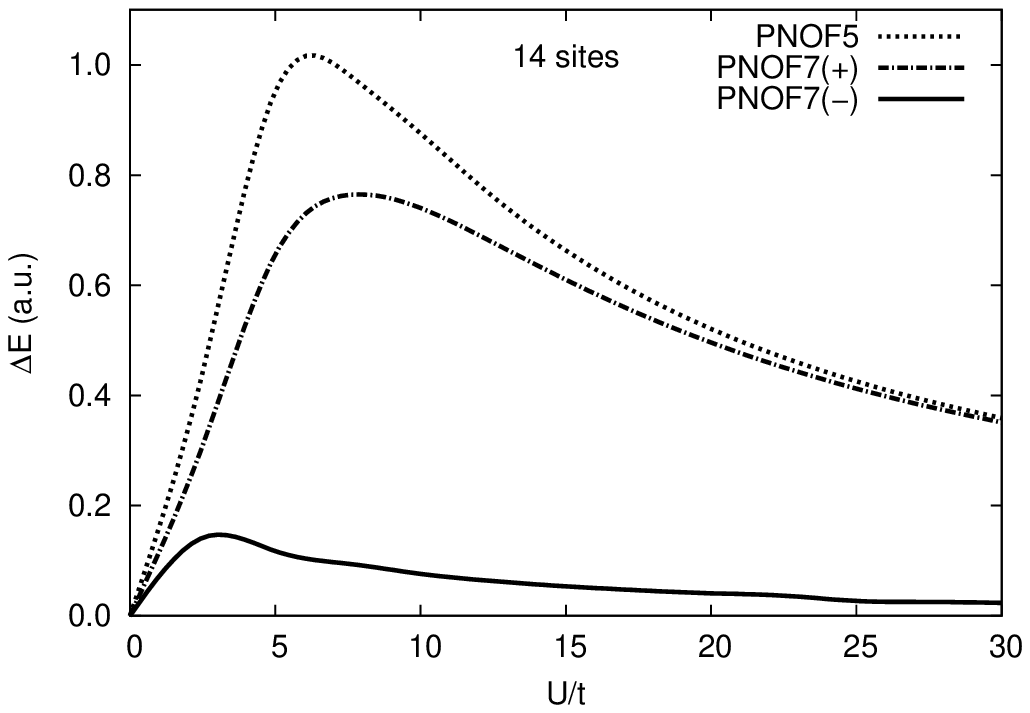}

\caption{Differences in $E$ values with respect to exact results vs. $U/t$
values for the 8, 10, 12, and 14 sites homogeneous 1D Hubbard model
with periodic boundary conditions at half-filling; obtained by using
PNOF5, PNOF7(+) and PNOF7(-).}

\label{homogeneous-hubbard-1}
\end{figure}

In Fig. \ref{homogeneous-hubbard-1}, we show the differences in $E$
values with respect to the exact diagonalization (ED) results ($\triangle E=E^{NOF}-E^{ED}$)
obtained for the 8, 10, 12 and 14 sites systems for a range of $U/t$
values in order to cover all correlation regimes \textcolor{black}{(exact
energy curves corresponding to these systems are included in the supplementary
material)}. Exact results are computed using a modified version of
the code developed by Knowles and Handy \cite{Knowles,Knowles2},
whereas results for NOF approximations have been computed using DoNOF
code developed by M. Piris and coworkers. 

First, we observe that energies obtained by using the independent-pair
model PNOF5, which is equivalent to a special case of an antisymmetrized
product of strongly orthogonal geminals \cite{Piris2013a}, systematically
underestimate the correlation effects for all $U/t$ values regardless
of the number of sites of the system. Therefore, it is mandatory to
consider the interactions between electron pairs to get an accurate
description for the Hubbard model.

Considering nonzero inter-pair interactions by introducing the $\Pi_{qp}^{\Phi}$
term given in Eq. (\ref{PNOF7}), the amount of electron correlation
recovered in the region $0<U/t<10$ is larger, but the behavior for
large $U/t$ values is rather similar to neglecting inter-pair interactions
if PNOF7(+) is used. As illustrated in Fig. \ref{homogeneous-hubbard-1},
this issue can be properly solved simply by considering a different
choice of the signs for $\Pi_{qp}^{\Phi}$. Thus, PNOF7(-) significantly
improves the performance of PNOF7(+) not only for any correlation
regime, but also for any size of the system. While PNOF7(+) produces
larger errors as the number of sites increases, the accuracy of PNOF7(-)
independent of the system size. In the region $U/t\gg1$, when the
on-site electronic repulsion gets larger, the PNOF7(-) curve attaches
to the exact curve giving an outstanding description of the asymptotic
behavior. Hence, this approximation is able to reproduce the antiferromagnetic
nature of the model in this region, in contrast to other approaches
based on electron-pair states, such as AP1roG \cite{Boguslawski2016},
which fail to describe weak orbital-pair correlations arising from
singly occupied states in the strong correlation limit.

\section{\label{sec:Hydrogen-rings}Hydrogen rings}

In accordance with our previous benchmarking \cite{mitxelena-hubbard,mitxelena-erratum}
and results showed in Fig. \ref{homogeneous-hubbard-1}, PNOF7(-)
is the best approximation within NOF theory to study systems described
by the Hubbard model. Within the limitations of the Hubbard model,
the lack of long-range inter-electron interactions may be one of the
most important. Therefore, in the following we focus on model systems
with strong static electron correlation in order to examine if the
conclusion obtained from the previous section still holds in presence
of long-range interaction effects.

\begin{figure}
\includegraphics[scale=0.21]{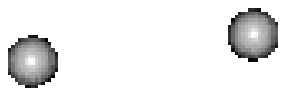}\qquad{}\includegraphics[scale=0.23]{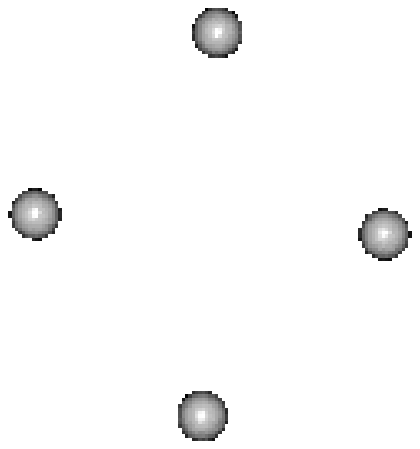}\qquad{}......\qquad{}\includegraphics[scale=0.19]{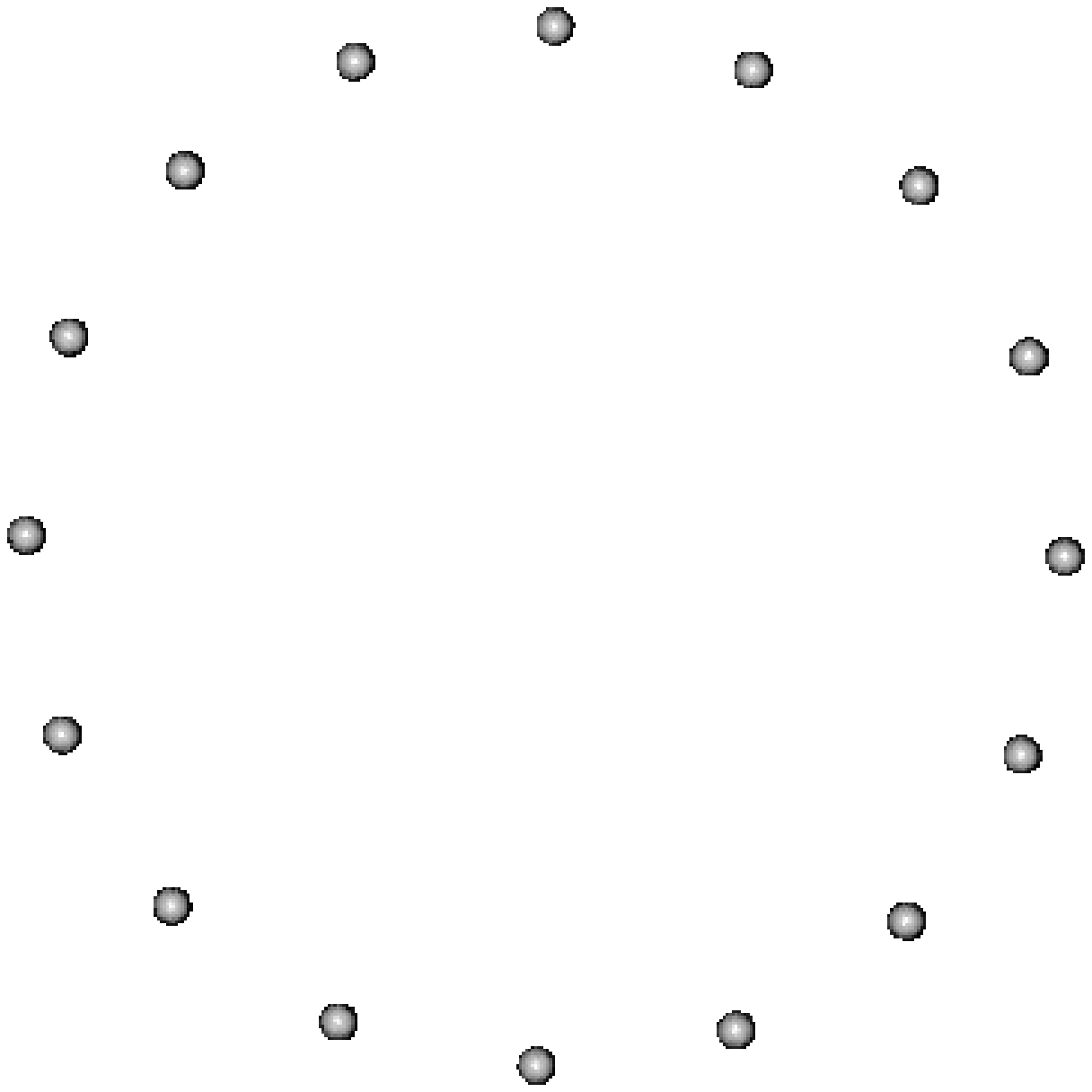}

\caption{2-dimensional polygon distribution of hydrogen atoms for 2, 4 and
16 atoms. Near-neighbor distance is fixed to $R_{H-H}=2.0$ \AA{}
for all the cases.}

\label{systems}
\end{figure}

Let us consider a ring of hydrogen atoms and vary the number of atoms
as done in the Hubbard model with the sites. We consider the non-relativistic
many-electron Hamiltonian to describe these systems, i.e.

\begin{equation}
\begin{array}{c}
{\displaystyle H=H_{nuc}+\sum_{\sigma}\sum_{pq}h_{pq}c_{p\sigma}^{\dagger}c_{q\sigma}}\\
\\
{\displaystyle +\frac{1}{2}\sum_{\sigma\tau}\sum_{pqrt}<pq|rt>c_{p\sigma}^{\dagger}c_{q\tau}^{\dagger}c_{t\tau}c_{r\sigma}},
\end{array}\label{eq:hamiltonian_nonREV}
\end{equation}

where the first term accounts for the inter-nuclear repulsion, the
second term includes both the kinetic energy and the nuclear repulsion,
and the last term introduces Coulombic repulsion between electrons.
Note that indices $p$, $q$, $r$, and $t$ run over spatial orbitals,
whereas $\tau$ and $\sigma$ run over spin functions. This model
may be the simplest example of strong electronic correlation in low
dimensions since a multi-reference method is required to get an accurate
description for $R_{H-H}=2.0$ \AA{} or larger bond distances, due
to the strong correlation near the equilibrium geometry and at the
dissociation limit \cite{Mazziotti-hydrogen-chain}. The employed
systems are illustrated in Fig. \ref{systems} for chains of 2, 4,
and 16 hydrogen atoms.

\begin{figure}
\includegraphics[scale=0.69]{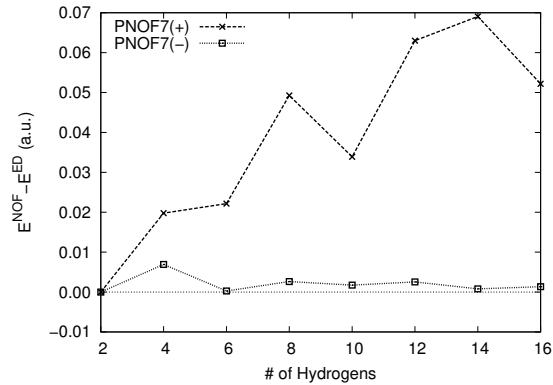}

\caption{Relative differences with respect to exact diagonalization (ED) energies
obtained by using PNOF7(+) and PNOF7(-) for the periodic chain of
hydrogens at $R_{H-H}=2.0$ \AA{} with varying size. Calculations
are performed using a minimal basis.}

\label{figure-HYDROGENS}
\end{figure}

In Fig. \ref{figure-HYDROGENS}, we show the relative energies obtained
by using PNOF7(+) and PNOF7(-) with respect to exact diagonalization,
increasing the chain of hydrogen atoms from 2 to 16 at an internuclear
distance of $R_{H-H}=2.0$ \AA . We use a minimal basis in all the
calculations. According to Fig. \ref{figure-HYDROGENS}, the results
obtained employing PNOF7(+) show the same drawbacks already displayed
for the Hubbard model. The relative errors shown by this approximation
get larger as the size of the chain increases, so PNOF7(+) is not
expected to give an accurate description of the electron correlation
in the presence of many inter-pair interactions. In contrast, when
we choose negatives, all the electron correlation functions $\Pi_{qp}^{\Phi}$
in Eq. (\ref{PNOF7}), the relative errors with respect to the results
of exact diagonalization remain equally small when increasing the
number of hydrogens, as shown in Fig. \ref{figure-HYDROGENS}. Note
that the accurate energy (\ref{2e}) is recovered for the two-electron
system by using either PNOF7(+) or PNOF7(-). The largest error obtained
by using PNOF7(-) is below $0.007$ Hartree (exact and approximated
energies are given in supplementary material), so the latter is notably
superior to PNOF7(+), and does not present any issue with the size
of the system.

\section{Closing remarks}

In this work, we have presented a novel approach to tackle the phase
dilemma in the context of natural orbital functional theory. The bottom-up
method employed by Piris to develop approximate functionals does not
require the use of a N-particle wavefunction and makes use of ensemble
N-representability conditions to get an explicit form of the functional.
Nevertheless, there is still an indeterminacy with respect to the
phase of the interaction between electron pairs with opposite spins.
We have shown that this indefiniteness must be studied carefully,
as it dramatically affects the performance of our approach.

For this purpose, we selected model systems with a strong static electron
correlation, such as the one-dimensional Hubbard model with periodic
boundary conditions and molecular hydrogen rings. Despite of their
simplicity, the Hubbard model and the hydrogen atom chain (located
at $R_{H-H}=2.0$ \AA ) present strong non-dynamic correlation effects,
and can be viewed as benchmarking systems for testing multi-reference
electronic structure methods. It has been demonstrated that the PNOF7
approach presented here captures the physics that appears in strongly
correlated systems. After an adequate choice of sign factors for the
inter-pair interactions, the so-called PNOF7 approximation gives a
quasi-exact description of non-dynamic correlation effects appearing
in these systems, even in the region of strong correlation.

According to the results shown throughout the paper, the proper selection
of phases amends the behavior of the functional when applying to large
systems. Thus, the performance of the here presented method, denoted
as PNOF7(-), does not deteriorate with the size of the system, so
the latter could be used to study strongly correlated systems beyond
small molecules, e.g. periodic polymers or heavy-element-containing
molecules.
\selectlanguage{american}%

\section*{Acknowledgements}

Financial support comes from Eusko Jaurlaritza (Ref. IT588-13) and
Ministerio de Economía y Competitividad (Ref. CTQ2015-67608-P). One
of us (I.M.) is grateful to Vice-Rectory for research of the UPV/EHU
for the PhD. grant \foreignlanguage{english}{(PIF//15/043}). \foreignlanguage{english}{M.R.M.
wants to thank the Spanish Ministry (MEC) for the PhD. grant (FPU-2013/00176).}

\selectlanguage{english}%
\smallskip{}

\end{document}